# Dynamic Property and Magnetic Nonpotentiality of Two Types of Confined Solar Flares

Xuchun Duan[1,2], Ting Li[2,3], and Qihang Jing[1]
[1] Sichuan University, Chengdu 610064, People's Republic of China
[2] National Astronomical Observatories, Chinese Academy of Sciences, Beijing 100101, People's Republic of China; liting@nao.cas.cn
[3] School of Astronomy and Space Science, University of Chinese Academy of Sciences, Beijing 100049, People's Republic of China

## Abstract

We analyze 152 large confined flares (GOES class $\geqslant$ M1.0 and $\leqslant 45°$ from disk center) during 2010−2019, and classify them into two types according to the criterion taken from the work of Li et al. "Type I" flares are characterized by slipping motions of flare loops and ribbons and a stable filament underlying the flare loops. "Type II" flares are associated with the failed eruptions of the filaments, which can be explained by the classical 2D flare model. A total of 59 flares are "Type I" flares (about 40%) and 93 events are "Type II" flares (about 60%). There are significant differences in distributions of the total unsigned magnetic flux ($\Phi_{AR}$) of active regions (ARs) producing the two types of confined flares, with "Type I" confined flares from ARs with a larger $\Phi_{AR}$ than "Type II." We calculate the mean shear angle $\Psi_{HFED}$ within the core of an AR prior to the flare onset, and find that it is slightly smaller for "Type I" flares than that for "Type II" events. The relative nonpotentiality parameter $\Psi_{HFED}/\Phi_{AR}$ has the best performance in distinguishing the two types of flares. About 73% of "Type I" confined flares have $\Psi_{HFED}/\Phi_{AR} < 1.0 \times 10^{-21}$ degree Mx$^{-1}$, and about 66% of "Type II" confined events have $\Psi_{HFED}/\Phi_{AR} \geqslant 1.0 \times 10^{-21}$ degree Mx$^{-1}$. We suggest that "Type I" confined flares cannot be explained by the standard flare model in 2D/3D, and the occurrence of multiple slipping magnetic reconnections within the complex magnetic systems probably leads to the observed flare.

*Unified Astronomy Thesaurus concepts:* Solar activity (1475); Solar active region magnetic fields (1975); Solar flares (1496); Solar filaments (1495); Solar coronal mass ejections (310)

*Supporting material:* animations

## 1. Introduction

Solar flares are one of the most energetic events in solar activity and as early as 1859, the observation of the solar flare was first recorded. For most large solar flares, they are usually associated with coronal mass ejections (CMEs). Solar flares and their associated CMEs are widely recognized as two performance forms of the same underlying physical process, and the flares associated with CMEs are named eruptive flares (Svestka & Cliver 1992). There are also some flares that are not associated with any CME, and these types of flares are called confined flares. It was revealed that flare–CME association rate increases with the flare intensity and decreases with the increasing total unsigned magnetic flux ($\Phi_{AR}$) of active regions (ARs) producing the flares (Yashiro et al. 2006; Li et al. 2020, 2021a).

Solar flares and CMEs are considered to originate from the rapid release of free magnetic energy through magnetic reconnection (Forbes 2000; Shibata & Magara 2011). Since the concept and preliminary theory of magnetic reconnection were put forward, Kopp & Pneuman (1976) proposed a model to elaborate the flare based mainly on the studies of Carmichael (1964), Sturrock (1966), and Hirayama (1974), which is called the CSHKP model. Before the flare onset, a magnetic flux rope (filament) located near the magnetic polarity inversion line (PIL) is bound by magnetic arcades across it. Due to the occurrence of magnetic reconnection, in particular of the tether-cutting type, the flux rope becomes unstable and starts to rise up. Then the magnetic reconnection occurs in the X-type magnetic structure under the rising flux rope and accelerates the flux rope. A series of extreme ultraviolet (EUV)/X-ray loops are formed across the PIL, and their footpoints correspond to Hα ribbons in the chromosphere.

Recent high-quality and rich observations showed that the classical 2D CSHKP model remains insufficient to explain numerous observational phenomena of solar flares. For instance, Li & Zhang (2014) found the apparent slipping motions of the flux rope's end along the flare ribbons. Dudík et al. (2014) showed the apparent slipping motion of flare loops along the developing flare ribbons during an eruptive X-class flare. Similar observations were also found in previous studies (Sun 2013; Li et al. 2015; Li & Zhang 2015; Dudík et al. 2016; Zemanová et al. 2019; Chen et al. 2019), which indicate that the energy release in solar flares is indeed an intrinsically 3D phenomenon; 3D extensions to the CSHKP model have been proposed to interpret the physical process of eruptive flares (Aulanier et al. 2012; Janvier et al. 2013), in which magnetic reconnection occurs along quasi-separatrix layers (QSLs; Priest & Démoulin 1995), and the continuous restructuring of field lines along the QSLs results in an apparent slipping motion of field line footpoints. Aulanier et al. (2006) simulated the process of fast slippage of magnetic field lines along QSLs in a confined flare in the absence of a flux rope. In recent years, the close correspondence between flare ribbons and the QSLs has been shown in many studies (Masson et al. 2009; Savcheva et al. 2015; Yang et al. 2015; Zhao et al. 2016), which provides strong evidence for QSL reconnection.

To reveal the key factors determining whether a flare is eruptive or not is an important question in flare studies. The constraining effect of the background magnetic field overlying

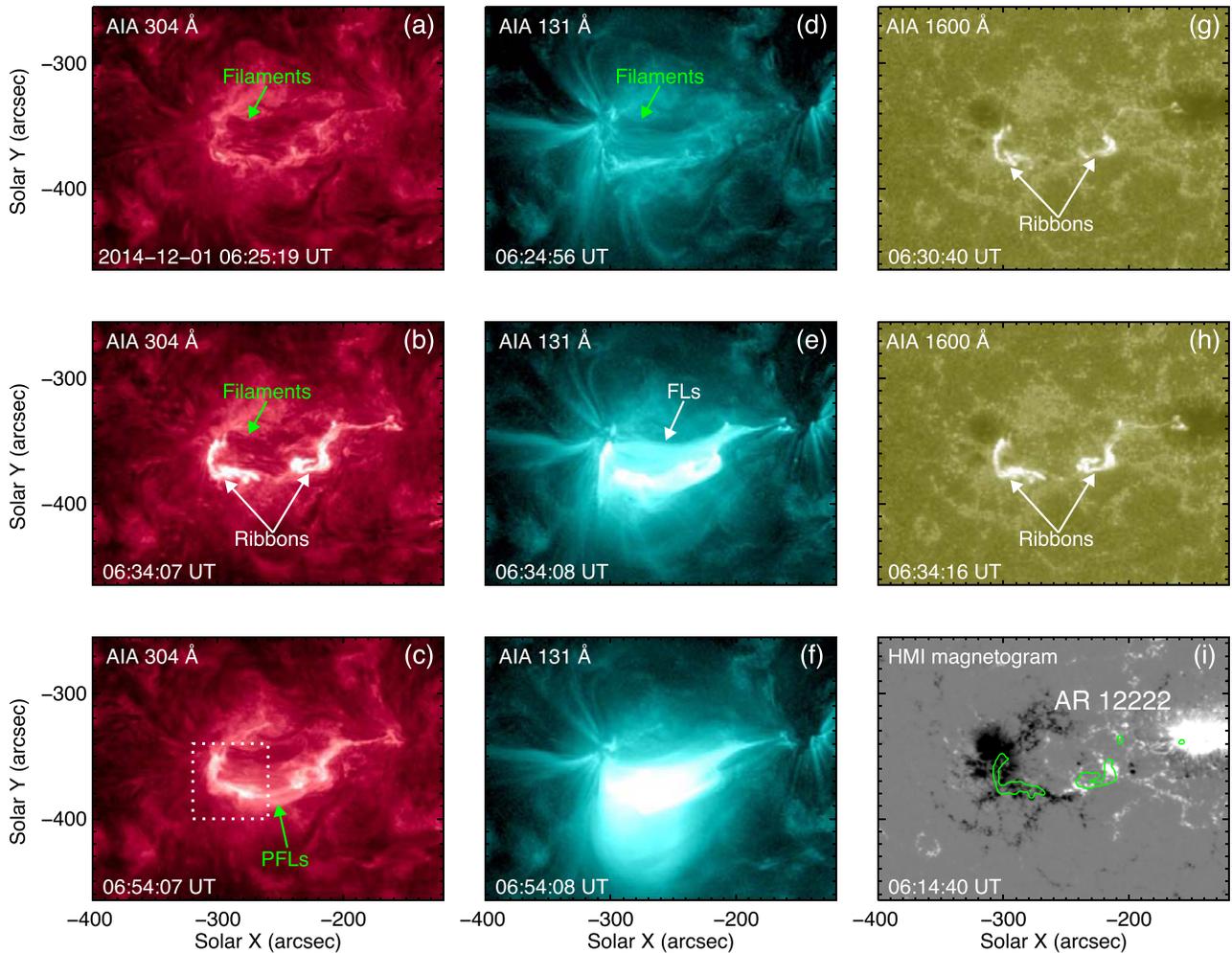

**Figure 1.** Appearance of the M1.8-class flare on 2014 December 1 in different (E)UV wavelengths and the corresponding HMI magnetogram. Filaments in panels (a)–(b) were noneruptive filaments in the flaring region in 304 Å. The postflare loops (PFLs) in panel (c) are postflare loops above the filaments. The white square in panel (c) denotes the field of view of Figure 2. The flare loops in panel (e) show the high-temperature flare loops in 131 Å. The green curves in panel (i) are the brightness contours of flare ribbons in the Atmospheric Imaging Assembly (AIA) 1600 Å image in panel (h). The animation of this figure includes AIA 304 and 131 Å images from 06:20 UT to 06:50 UT.

(An animation of this figure is available.)

the flaring region is thought to be a key factor (Wang & Zhang 2007; Yang et al. 2014; Wang et al. 2017; Baumgartner et al. 2018). Wang & Zhang (2007) found that confined events occur closer to the magnetic center of an AR and eruptive events tend to occur close to the AR edge, implying that the strong external field overlying the AR core is probably the main reason for the confinement. Besides the confinement of overlying magnetic fields, the nonpotentiality of ARs producing the flares is another aspect to determine the eruptive character of solar flares (Nindos & Andrews 2004; Sun et al. 2015; Cui et al. 2018; Gupta et al. 2021; Kazachenko et al. 2022). The nonpotentiality of ARs reflects the degree that the photospheric magnetic field of an AR deviates from its potential field, which can be represented by many magnetic parameters, such as magnetic helicity, magnetic twist, magnetic energy, etc. Li et al. (2022) revealed that two relative nonpotentiality parameters $\alpha/\Phi_{AR}$ and $\Psi/\Phi_{AR}$ within the AR core ($\alpha$ is the mean characteristic twist parameter and $\Psi$ is the mean shear angle) for confined flares are significantly smaller than those for eruptive flares.

The dynamic property of confined flares is important to understand the confining mechanism of solar flares. It was considered that confined flares are usually associated with the failed eruption of the filament or flux rope (Ji et al. 2003; Joshi et al. 2019; Huang et al. 2020; Zhang et al. 2022). Recently, Li et al. (2019) analyzed 18 confined flares ⩾ M5.0-class and found that two types of confined flares are present. "Type I" confined flares are characterized by slipping reconnection, a stable filament, and strongly sheared postflare loops (PFLs). "Type II" flares are associated with the failed eruption of a filament, which is confined by a strong strapping field. Following the study of Li et al. (2019), here we enlarge the flare sample including 152 confined flares ⩾ M1.0-class and compare the nonpotentiality of ARs producing the two types of confined flares. The rest of the paper is organized as follows. In Sections 2 and 3, we describe the data analysis and show the examples and statistical results for the two types of confined flares, respectively. Finally, we summarize our findings and discuss their implications in Section 4.

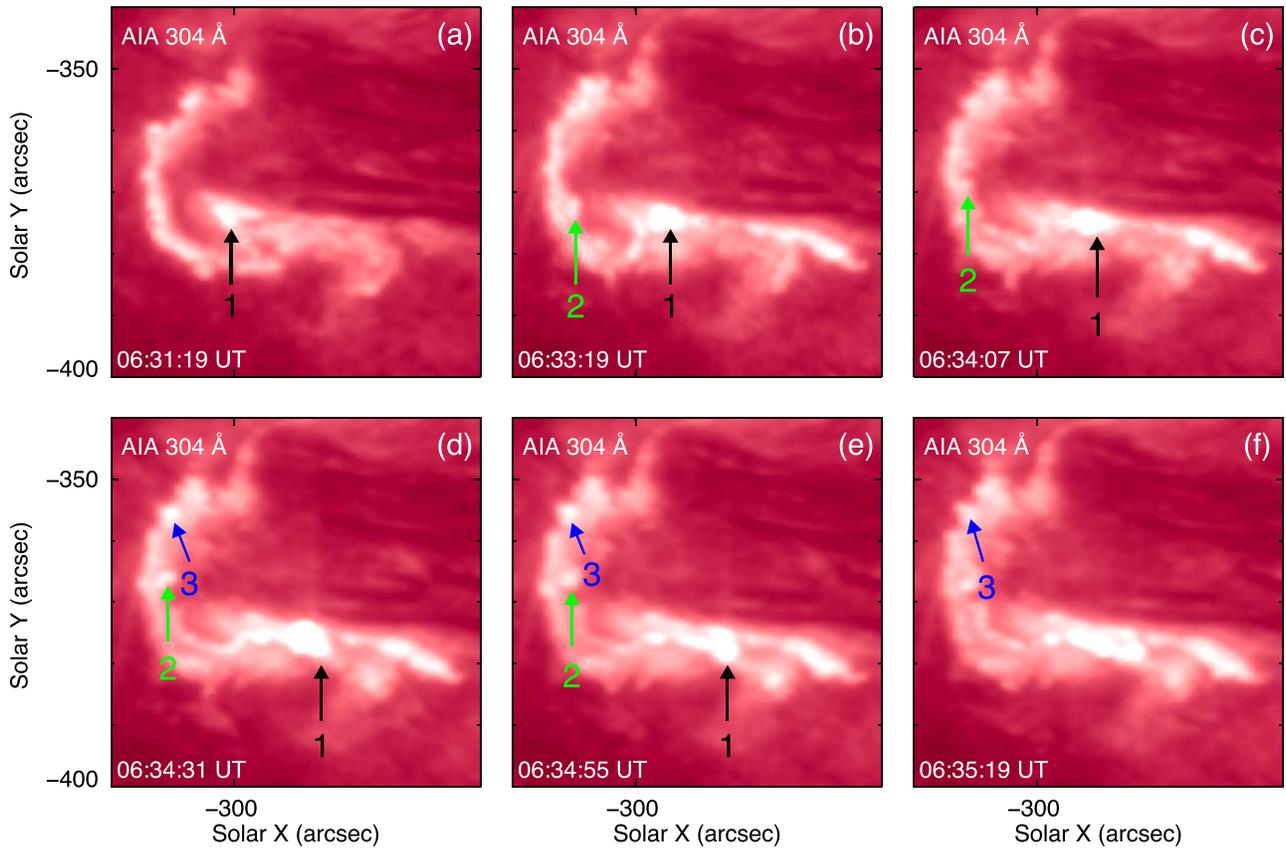

**Figure 2.** Time series of 304 Å images showing the slippage of traced bright knots ("1" to "3") within the flare ribbon. Bright knots "2" and "3" slipped toward the north and knot "1" slipped toward the west. The animation of this figure includes Atmospheric Imaging Assembly (AIA) 304 Å images from 06:25 UT to 06:40 UT. (An animation of this figure is available.)

## 2. Observations and Data Analysis

The data set analyzed in this study involves 152 M-class confined flares[4] during the period of 2010 June to 2019 June (Li et al. 2020). For each event, we check the (E)UV observations from the Atmospheric Imaging Assembly (AIA; Lemen et al. 2012) on board the Solar Dynamics Observatory (SDO; Pesnell et al. 2012) to make a classification. In the channels of AIA/304 and 171 Å, the dynamic evolution of the filament during the flare can be clearly discerned. From the AIA/1600 and 131 Å observations, we can see if the slipping motions of flare ribbons and high-temperature loops occur during the confined flare. Thus the four channels of AIA 1600, 304, 171, and 131 Å are mainly consulted in this work. We also use the line-of-sight magnetic field data from the Helioseismic and Magnetic Imager (HMI; Scherrer et al. 2012) on board the SDO to compare the ribbon locations with the photospheric magnetograms.

In order to calculate the nonpotentiality of ARs producing the solar flares, we also use the vector magnetograms from Space-Weather HMI AR Patches (SHARP; Bobra et al. 2014) of the SDO, which are remapped using a cylindrical equal area projection with a pixel size of ~0″.5 and presented as $(B_r, B_\theta, B_\phi)$ in heliocentric spherical coordinates corresponding to $(B_z, -B_y, B_x)$ in heliographic coordinates (Sun et al. 2013). Two nonpotential parameters are calculated for each event, including the photospheric magnetic free energy density $\rho_{\text{free}}$ and mean shear angle $\Psi$; $\rho_{\text{free}}$ describes the distribution of the magnetic energy in excess of the minimum energy attributed to the potential field, which is calculated as

$$\rho_{\text{free}} = \frac{|\boldsymbol{B}_o - \boldsymbol{B}_p|^2}{8\pi}, \quad (1)$$

where $\boldsymbol{B}_o$ and $\boldsymbol{B}_p$ are the observed and the potential magnetic fields, respectively.

For the distribution of $\rho_{\text{free}}$, we select the region with $\rho_{\text{free}} > 4.0 \times 10^4$ erg cm$^{-3}$ (high free energy density region—HFED region; Chen & Wang 2012; Li et al. 2022) as a proxy for the AR core region. Then we measure the mean shear angle $\Psi_{\text{HFED}}$ within the HFED region. Magnetic shear is defined as the angle between the horizontal components of the observed magnetic field and a modeled potential magnetic field based on the photospheric $B_z$ map, which is given by

$$\Psi = \arccos \frac{\boldsymbol{B}_o \cdot \boldsymbol{B}_p}{|B_o B_p|}. \quad (2)$$

## 3. Results

For all the 152 confined flares, we divided them into two types based on the following criterion, consistent with the definition in Li et al. (2019).

During "Type I" confined flares, the slipping motions of the flare loops or flare ribbons can be clearly seen. The filament is stable and does not seem to be disturbed during the flare. In addition, the low-temperature PFLs observed in the 171 and 304 Å channels are located above the stable filament.

---
[4] The data set used (Version 1) is available from China-VO: doi:10.12149/101031.

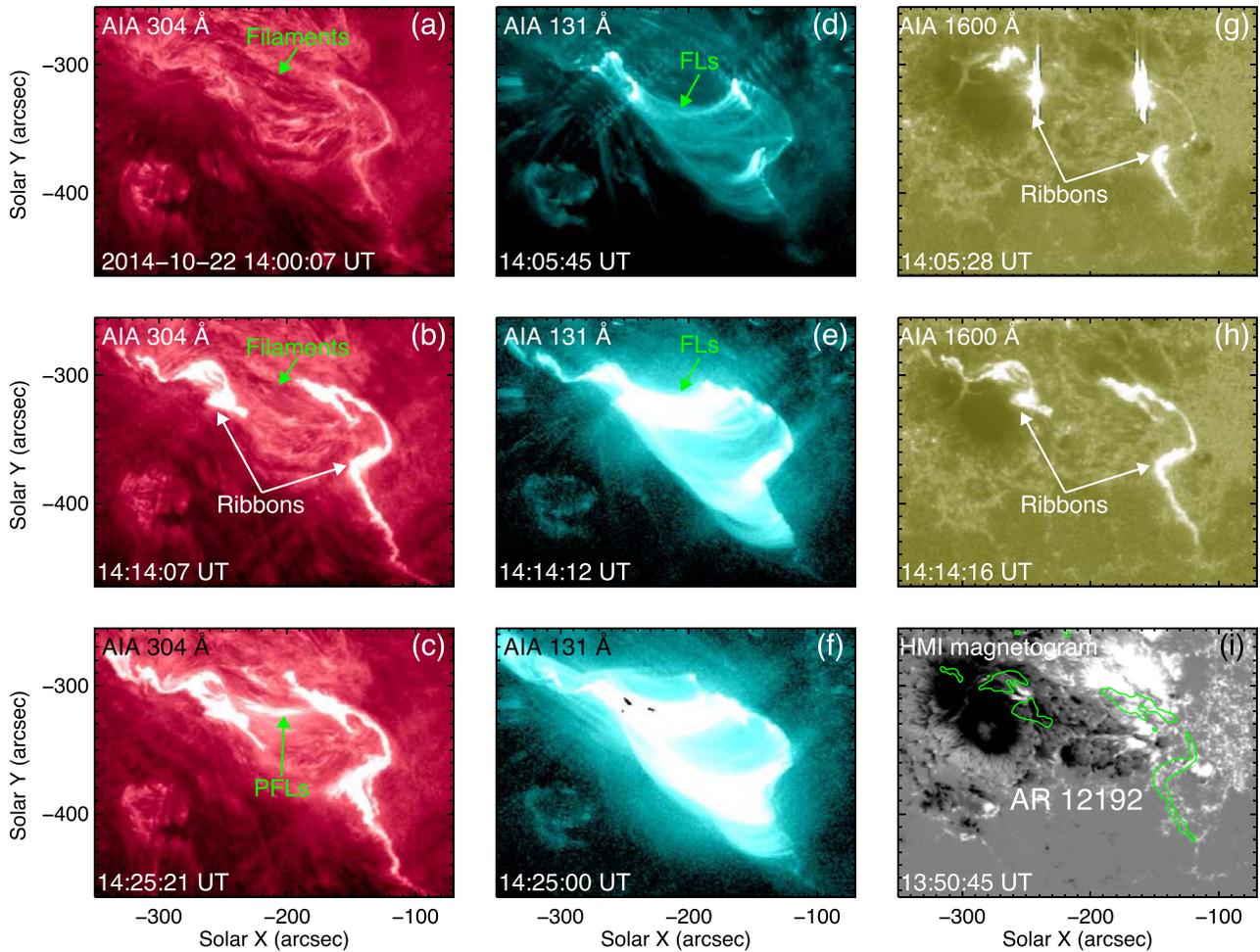

**Figure 3.** Similar to Figure 1, but for the X1.6-class flare on 2014 October 22. The animation of this figure includes Atmospheric Imaging Assembly (AIA) 304 and 131 Å images from 14:00 UT to 14:30 UT.
(An animation of this figure is available.)

The main characteristic of "Type II" confined flares is the failed eruption of the core filament. The filament initially rises up during the flare and then quickly slows down. The overlying large-scale arcades are heated and can be observed at the high-temperature wavelength (131 Å). In other words, "Type II" flares are consistent with the classical 2D flare model.

Among the 152 confined events, a total of 59 flares are "Type I" flares (about 40%) and 93 events are "Type II" flares (about 60%). In order to show the dynamic property of the two types of confined flares, two events from each type of confined flares are selected as examples to analyze in detail. Based on the vector magnetograms from the SDO/HMI, we calculate the mean shear angle $\Psi_{\mathrm{HFED}}$ within HFED region for each event prior to the flare onset. Here we create a new data set, ConfinedflareDB,[5] and describe the classification of the event ("Type I" or "Type II"), the total unsigned magnetic flux $\Phi_{\mathrm{AR}}$ of ARs and the mean shear angle $\Psi_{\mathrm{HFED}}$.

_________________________
[5] The new data set (Version 1) is available from China-VO: doi:10.12149/101104.

### 3.1. "Type I": the M1.8-class Flare on 2014 December 1 and the X1.6-class Flare on 2014 October 22

One selected event of "Type I" confined flares is the M1.8-class flare occurring in AR 12222 on 2014 December 1. The GOES soft X-ray (SXR) 1–8 Å flux showed that the M1.8-class flare started at 06:26 UT and peaked at 06:41 UT. From the observations of 304 Å, the short filaments are present connecting the positive and negative-polarity magnetic fields before the flare onset (Figures 1(a) and (i)). After the flare started, the filaments did not show any rise phase and remained stabilized (Figure 1(b)). At the decay phase of the flare, low-temperature PFLs were observed at 304 Å, which are located above the noneruptive filaments (Figure 1(c)). In the 131 Å channel, the flare loops (FLs) were heated and illuminated when the flare started (Figures 1(d)–(f)). As the flare evolved, more loop bundles appeared and displayed apparent slipping motions along ribbons (see Animation 1, Figure 1). Two brightened ribbons can be clearly observed in the 1600 Å images (Figures 1(g)–(h)), with one ribbon at the southeast of the leading positive-polarity sunspot and the other one at the southwest of the following negative-polarity sunspot. SDO/HMI magnetograms show that the AR was gradually decaying on 2014 December 1 (Figure 1(i)).

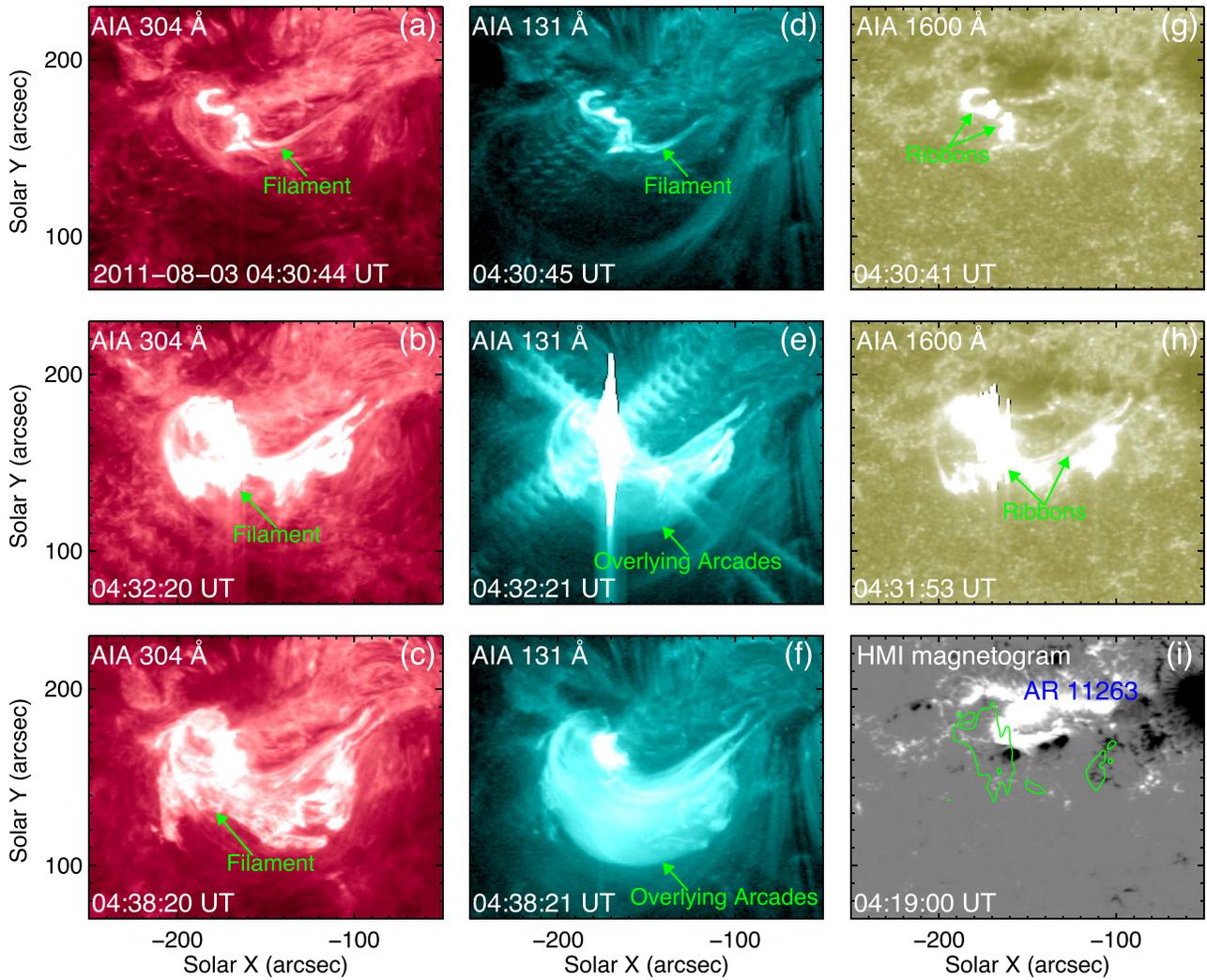

**Figure 4.** Similar to Figure 1, but for the M1.7-class flare on 2011 August 3. The filament shows a failed eruption process. The animation of this figure includes Atmospheric Imaging Assembly (AIA) 304 and 131 Å images from 04:20 UT to 04:48 UT.
(An animation of this figure is available.)

The apparent slipping motion of the flare ribbons can be seen in the 304 Å, which is displayed in Figure 2 to show the details of the dynamic evolution of the ribbon substructures. In Figure 2, we can see that the east ribbon was composed of bright knots, which showed slipping motions clearly (see Animation 2, Figure 2). We tracked and labeled three individual bright knots within the ribbon as "1" to "3". These three bright knots move mainly in two directions. The bright knot "1" slid westward along the straight-line part of the ribbon. Bright knot "2" and bright knot "3" slid toward the north of the ribbon. The bidirectional slipping motion of ribbon substructures implies the occurrence of slipping magnetic reconnection between different magnetic systems (Li & Zhang 2015; Dudík et al. 2016).

Another event we selected is the X1.6−class event in AR 12192 near the solar disk center (S14°, E13°) on 2014 October 22 as an example of "Type I" confined flares. The GOES SXR 1−8 Å flux showed that the X1.6−class flare started at 14:02 UT and peaked at 14:28 UT. The 304 Å observations showed that the filament systems (Figure 3(a)) are present prior to the onset of the flare. There was no obvious rising stage for the filaments, and during the flare process, any failed eruptions are not associated with them (Figure 3(b); see Animation 3, Figure 3). At about 14:25 UT, the PFLs appeared at 304 Å connecting two flare ribbons (Figure 3(c)), indicating the cooling process of newly formed high-temperature loops through magnetic reconnections. In the channel 131 Å, the FLs displayed complex structures (Figures 3(d)–(f)), such as the loops that are sheared with each other. With the flare evolution, the loop bundles were more abundant and exhibited the slipping motions along flare ribbons. The flare is composed of two ribbons as discerned in 1600 Å (Figures 3(g)–(h)), with one short ribbon located nearby the sunspot and the other long ribbon extended southward to the facula region (Figure 3(i)).

*3.2. "Type II": the M1.7-class Flare on 2011 August 3 and the M7.6-class Flare on 2015 September 28*

One selected "Type II" confined flare was generated in AR 11263 (N17°, E08°) on 2011 August 3. The flare started at 04:29 UT and peaked at 04:32 UT from the GOES SXR 1−8 Å flux. It can be seen from the 304 Å images at 04:30 UT, the shape of filament was like the letter of $\varepsilon$ (Figures 4(a) and (d)). Then the filament started to erupt and showed an expansion process (Figure 4(b); see Animation 4, Figure 4). From about 04:38 UT, the filament material gradually drained back along its two legs to the solar surface and finally the brightening of the material faded away (Figure 4(c)). From the 131 Å images, it can be seen that large-scale EUV loops were present over the flaring region (Figure 4(e)). Associated with the eruption of the filament, these large-scale EUV loops were disturbed and pushed outward. We suggest that the large-scale EUV loops probably constrained the eruption of the filament, which resulted in the failed eruption of the filament. At the decay phase of the flare, brightened PFLs can be clearly observed in the 131 Å channel (Figure 4(f)), indicative of the newly formed loops through magnetic reconnection. In the early stage of the flare, two ribbons can be observed from the 1600 Å images (Figure 4(g)). At 04:31 UT, the ribbons became longer and more extensional (Figure 4(h)). It seems that the east ribbon was anchored in the southeast of the following positive-polarity sunspot and the west ribbon was located at the negative-polarity magnetic fields (Figure 4(i)).

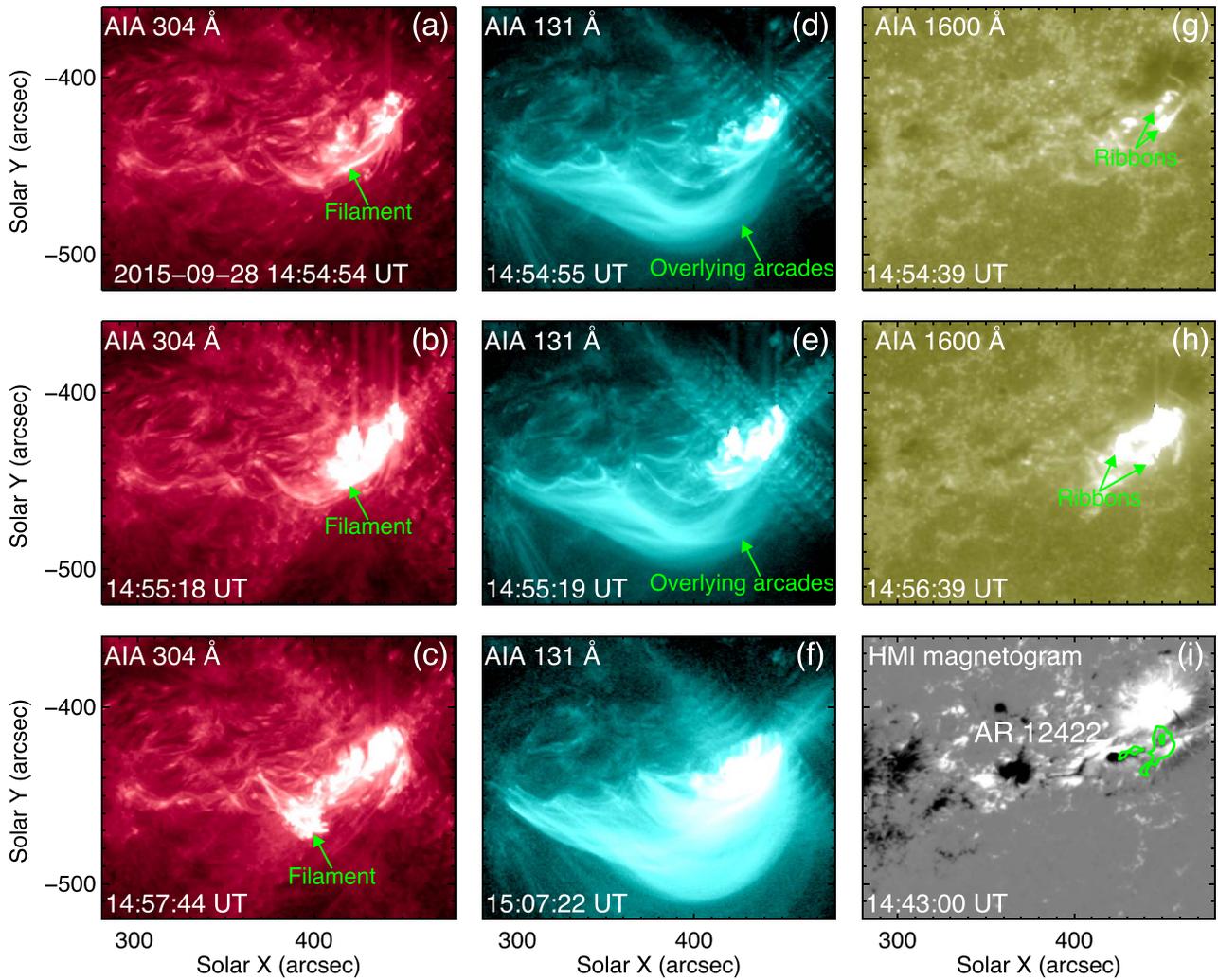

**Figure 5.** Similar to Figure 1, but for the M7.6-class flare on 2015 September 28. The filament shows a failed eruption process. The animation of this figure includes Atmospheric Imaging Assembly (AIA) 304 and 131 Å images from 14:40 UT to 15:10 UT.
(An animation of this figure is available.)

Another selected "Type I" confined flare initiated at 14:53 UT and peaked at 14:58 UT in AR 12422 (S20°, W16°) on 2015 September 28 from the GOES SXR 1−8 Å flux. At about 14:54 UT, the filament was illuminated and started to erupt (Figures 5(a) and (d); see Animation 5, Figure 5). Meanwhile, the high-temperature 131 Å images showed the presence of large-scale EUV loops overlying the erupted filament. Then more filament material and large-scale EUV loops were illuminated (Figures 5(b) and (e)). At about 14:57 UT, most filament material moved toward the east and then fell down to the solar surface (Figure 5(c)). The filament underwent a failed eruption with no material and magnetic structure going into the interplanetary space. Associated with the failed eruption of the filament, the large-scale EUV loops displayed a gradual expansion process, which ceased while reaching a certain height (Figures 5(e)−(f)). In the 1600 Å, it can be observed that there are two main ribbons at the northwest part of the filament (Figures 5(g)−(h)), which are both located at the south of the sunspot (Figure 5(i)).

### 3.3. Magnetic Flux of ARs and Shear Angles for "Type I" and "Type II" Confined Flares

Figure 6 shows the maps of photospheric magnetic free energy density $\rho_{\rm free}$ and magnetic shear angle $\Psi$ for the two "Type I" and "Type II" events shown in Figures 1–5. It can be seen that the maps of $\rho_{\rm free}$ and $\Psi$ exhibit overall similar distributions, with their large values around the PILs of ARs. For the "Type I" M1.8-class flare on 2014 December 1, the $\rho_{\rm free}$ and $\Psi$ are both small (Figures 6(a)–(b)). The ribbon area (green contours in panel (b)) does not correspond to the high $\Psi$ region. We also calculate the mean shear angle $\Psi_{\rm HFED}$ within the areas of $\rho_{\rm free} > 4.0 \times 10^4$ erg cm$^{-3}$ and it is only 24° for this event. Compared with the first event, the second "Type I" flare has a larger $\rho_{\rm free}$ and $\Psi$ (Figures 6(c)–(d)), with $\Psi_{\rm HFED}$ about 57°. For the two "Type II" events, the values of $\rho_{\rm free}$ and $\Psi$ are evidently higher than the two "Type I" flares (Figures 6(e)–(h)). The $\Psi_{\rm HFED}$ for the M1.7-class flare on 2011 August 3 and M7.6-class flare on 2015 September 28° are 72° and 67°, respectively. It can also be noticed that the flare ribbons in the two events overlay part of the high $\Psi$ area. This implies that the energy release of solar flares occurs in the region with strongly sheared magnetic fields.

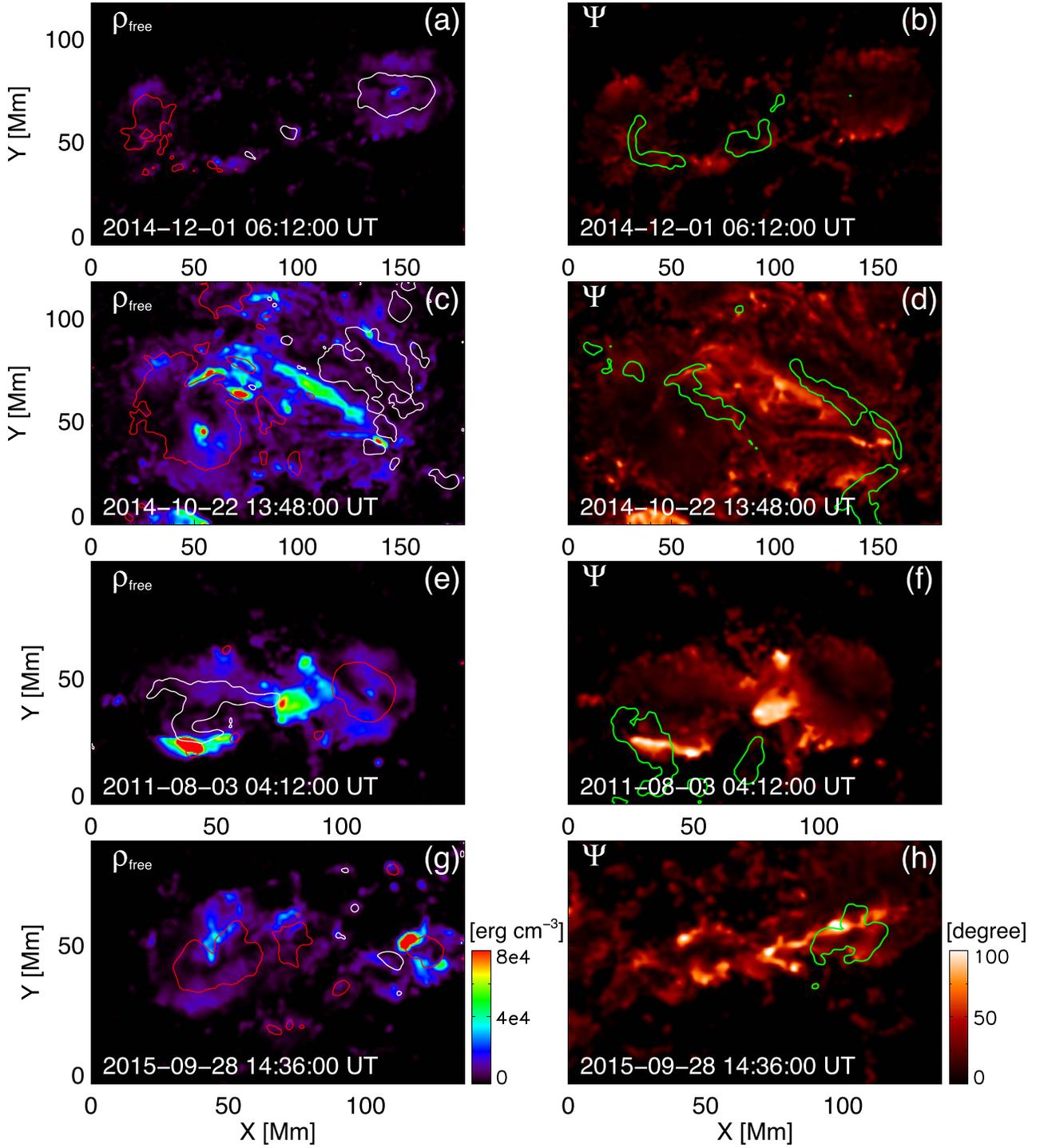

**Figure 6.** Maps of photospheric free magnetic energy density $\rho_{\rm free}$ and magnetic shear angle $\Psi$ of four examples (from top to bottom: "Type I" M1.8-class flare on 2014 December 1, "Type I" X1.6-class flare on 2014 October 22, "Type II" M1.7-class flare on 2011 August 3, and "Type II" M7.6-class flare on 2015 September 28). The white and red contours in the left column are the magnetic fields $B_z$ at ±800 G levels. The green contours in the right column outline the locations of flare ribbons observed in 1600 Å. Mean shear angle $\Psi_{\rm HFED}$ in Figure 7 is calculated within the areas of $\rho_{\rm free} > 4.0 \times 10^4$ erg cm$^{-3}$.

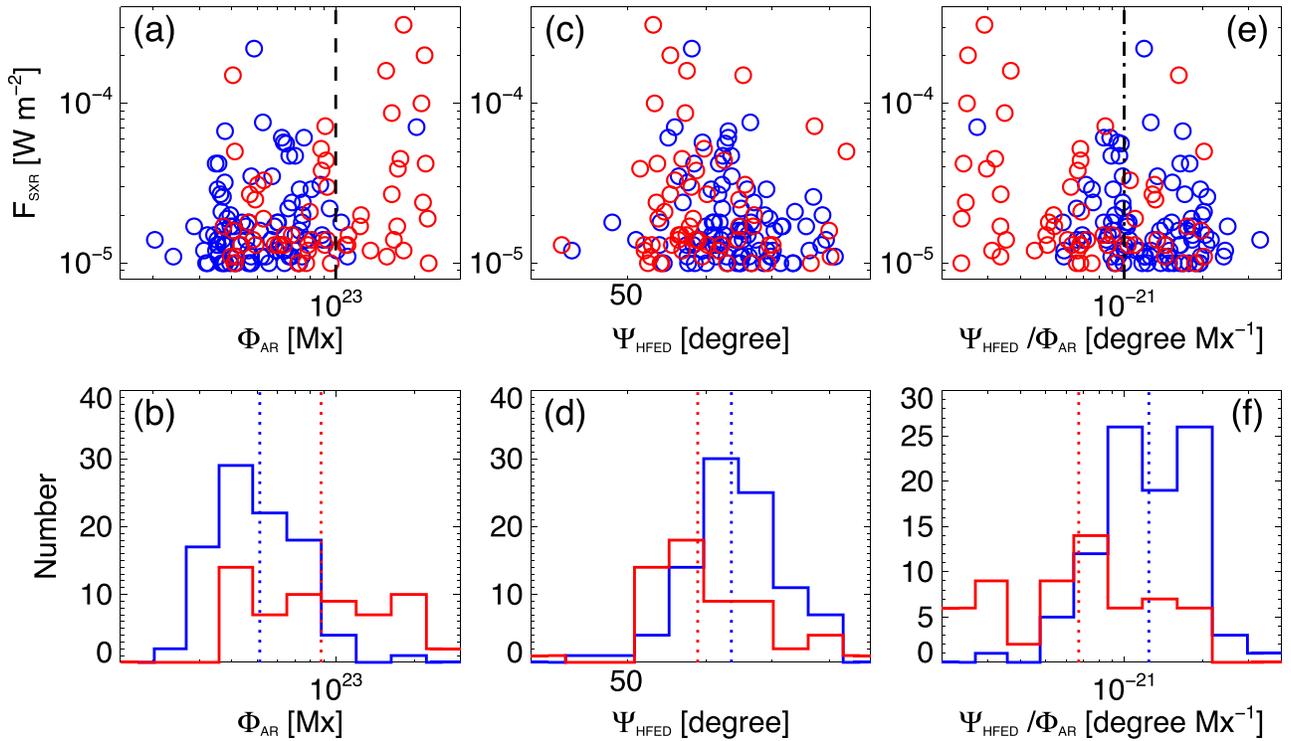

**Figure 7.** Panels (a)–(b): scatter plot of flare peak X-ray flux $F_{SXR}$ vs. unsigned AR magnetic flux $\Phi_{AR}$ and histograms of $\Phi_{AR}$. Red (blue) circles and lines are the "Type I" ("Type II") flares. The vertical dashed line in panel (a) corresponds to $\Phi$ of $1.0 \times 10^{23}$ Mx. The dotted vertical lines indicate the means of the log values. Panels (c)–(d): scatter plot of flare peak X-ray flux $F_{SXR}$ vs. mean shear angle $\Psi_{HFED}$ and histograms of $\Psi_{HFED}$. Panels (e)–(f): scatter plot of flare peak X-ray flux $F_{SXR}$ vs. relative mean shear angle $\Psi_{HFED}/\Phi_{AR}$ and histograms of $\Psi_{HFED}/\Phi_{AR}$. The vertical dashed–dotted line in panel (e) corresponds to $\Psi_{HFED}/\Phi_{AR}$ of $1.0 \times 10^{-21}$ degree Mx$^{-1}$.

We calculated the total unsigned magnetic flux $\Phi_{AR}$ and the mean shear angle $\Psi_{HFED}$ within the areas of high $\rho_{free}$ for 59 "Type I" and 93 "Type II" confined flares. The statistical results are shown in Figure 7. As seen from the scatter plot of the flare peak X-ray flux $F_{SXR}$ versus $\Phi_{AR}$ (Figure 7(a)), a majority of confined flares (22 out of 26) belong to "Type I" confined flares from ARs with a large $\Phi_{AR} > 1.0 \times 10^{23}$ Mx (indicated by the vertical dashed line). The histograms for the two types of confined flares show that there are significant differences in distributions of $\Phi_{AR}$ between "Type I" and "Type II" confined flares (Figure 7(b)). "Type I" events tend to have a larger $\Phi_{AR}$ than "Type II" flares. The averages of the log values of $\Phi_{AR}$ (indicated by vertical dotted lines) are $8.8 \times 10^{22}$ Mx and $5.1 \times 10^{22}$ Mx for "Type I" and "Type II" confined flares, respectively. The standard deviations of $\Phi_{AR}$ for "Type I" and "Type II" are $5.9 \times 10^{22}$ Mx and $2.47 \times 10^{22}$ Mx, respectively. For the parameter $\Psi_{HFED}$, there is only a small difference in distributions for the two types of confined flares (Figures 7(c)–(d)). The average of the log values of $\Psi_{HFED}$ is 59° for "Type I", slightly smaller than that for "Type II" confined flares (about 64°). If we consider the relative parameter $\Psi_{HFED}/\Phi_{AR}$, the differences between "Type I" and "Type II" confined flares are more evident (Figure 7(e)). About 73% (43 of 59) of "Type I" confined flares have $\Psi_{HFED}/\Phi_{AR} < 1.0 \times 10^{-21}$ degree Mx$^{-1}$, and ∼66% (61 of 93) of "Type II" confined events have $\Psi_{HFED}/\Phi_{AR} \geqslant 1.0 \times 10^{-21}$ degree Mx$^{-1}$ (black dashed–dotted line in Figure 7(e)). The averages of the log values of $\Psi_{HFED}/\Phi_{AR}$ (indicated by vertical dotted lines in Figure 7(f)) are $6.7 \times 10^{-22}$ degree Mx$^{-1}$ and $1.2 \times 10^{-21}$ degree Mx$^{-1}$ for "Type I" and "Type II" confined events, respectively. "Type II" flares are approximately normally distributed in all three quantities analyzed ($\Phi_{AR}$, $\Psi_{HFED}$, and $\Psi_{HFED}/\Phi_{AR}$). However, "Type I" flares show large deviations from normal distributions in $\Phi_{AR}$ and $\Psi_{HFED}/\Phi_{AR}$.

### 3.4. Flare Reconnection Flux and Flare Duration for "Type I" and "Type II" Confined Flares

We also analyze the reconnection flux swept by flare ribbons $\Phi_{RIB}$ (from a RibbonDB database in Kazachenko et al. 2017) and the FWHM duration of the flares $\tau_{FWHM}$ for the two types of confined flares. It was shown that flare peak X-ray flux $F_{SXR}$ correlates with flare reconnection flux $\Phi_{RIB}$ at a moderate rank order correlation coefficient $r_s$ of 0.42–0.43 for all the events: "Type I" and "Type II" flares (Figure 8(a)). Differently, the fitting slope $\alpha$ for "Type I" flares is evidently higher than that of "Type II" flares. Similar to the parameter $\Phi_{AR}$, "Type I" confined flares have a larger reconnection flux $\Phi_{RIB}$ than "Type II" events (Figure 8(b)). The averages of the log values of $\Phi_{RIB}$ (indicated by vertical dotted lines) are $5.3 \times 10^{21}$ Mx and $3.5 \times 10^{21}$ Mx for "Type I" and "Type II" cases, respectively. The standard deviations of $\Phi_{RIB}$ for "Type I" and "Type II" are $5.5 \times 10^{21}$ Mx and $2.37 \times 10^{21}$ Mx, respectively. For "Type I" confined flares, the reconnection flux $\Phi_{RIB}$ and the FWHM duration of the flares $\tau_{FWHM}$ (Figure 8(c)) show a moderate correlation with a rank order correlation coefficient $r_s$ of 0.46. However, there is no obvious positive correlation between the two parameters for "Type II" flares. Similarly, there is a significant difference in distributions of $\tau_{FWHM}$ between the two types of flares (Figure 8(d)). The averages of the log values of $\tau_{FWHM}$ are 920 s and 550 s for "Type I" and "Type II" events, respectively.

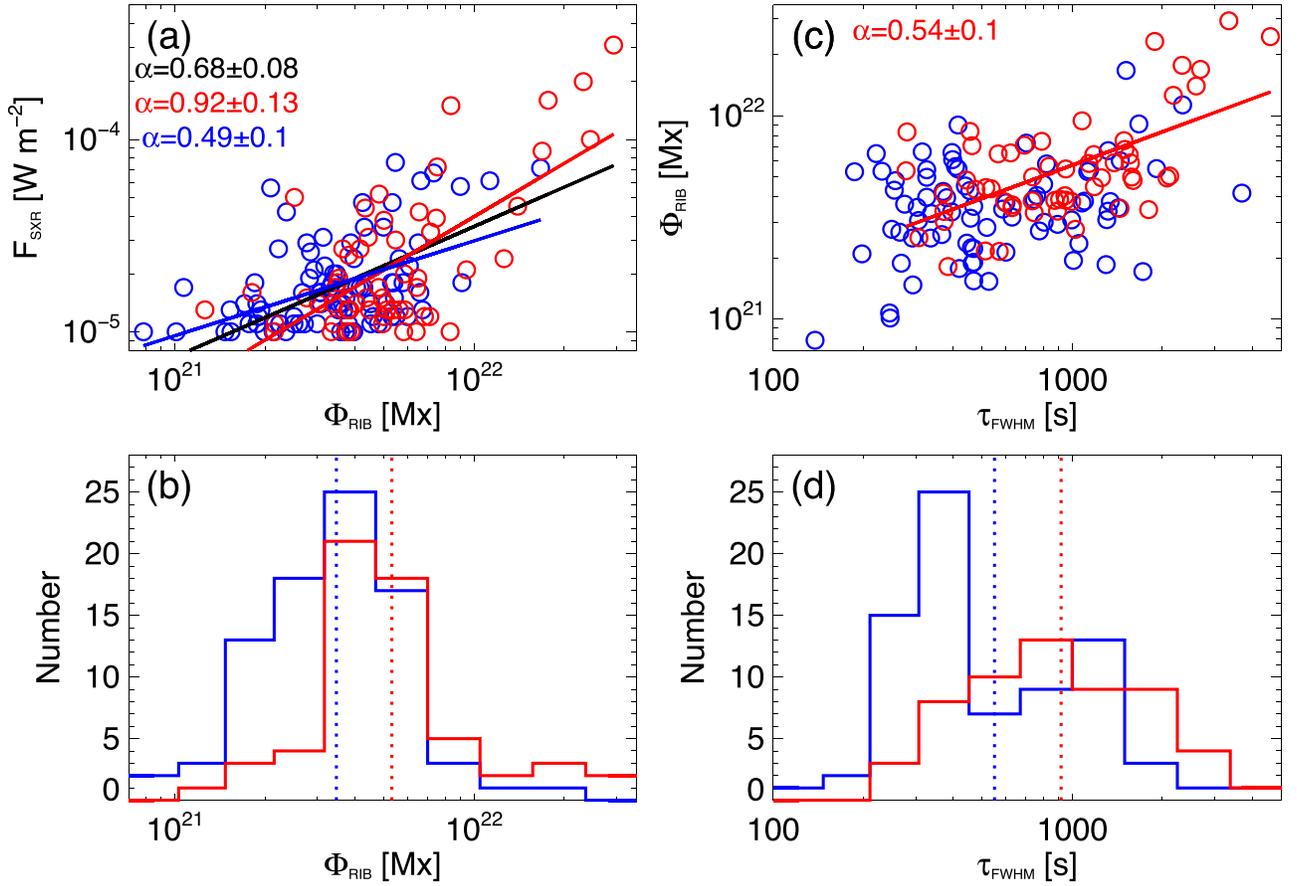

**Figure 8.** Panels (a)–(b): scatter plot of flare peak X-ray flux $F_{SXR}$ vs. flare reconnection flux $\Phi_{RIB}$ and histograms of $\Phi_{RIB}$. Panels (c)–(d): scatter plot of flare reconnection flux $\Phi_{RIB}$ vs. flare duration $\tau_{FWHM}$ and histograms of $\tau_{FWHM}$. Red (blue) circles and lines are the "Type I" ("Type II") flares. Red, blue, and black straight lines in panels (a) and (c) show the results of linear fitting for "Type I," "Type II," and total events, respectively. The slopes $\alpha$ are shown at the panel top. The dotted vertical lines in panels (b) and (d) indicate the means of the log values.

## 4. Summary and Discussion

In this study, we analyzed 152 confined flares (GOES class $\geqslant$ M1.0 and $\leqslant 45°$ from disk center) that occurred between 2010 June until 2019 June, and classified them into two types based on their different dynamic properties. "Type I" flares are characterized by slipping motions of flare loops and ribbons and a stable filament underlying the flare loops. "Type II" flares are associated with the failed eruptions of the filaments, which can be explained by the classical 2D flare model. A total of 59 flares are "Type I" flares (about 40%) and 93 events are "Type II" flares (about 60%). There are significant differences in distributions of $\Phi_{AR}$ between "Type I" and "Type II" confined flares, with "Type I" confined flares from ARs with a larger $\Phi_{AR}$ than "Type II." The nonpotentiality parameter $\Psi_{HFED}$ of "Type I" flares is slightly smaller than that of "Type II." However, the relative nonpotentiality parameter $\Psi_{HFED}/\Phi_{AR}$ has the best performance in distinguishing the two types of flares. About 73% (43 of 59) of "Type I" confined flares have $\Psi_{HFED}/\Phi_{AR} < 1.0 \times 10^{-21}$ degree $Mx^{-1}$, and $\sim$66% (61 of 93) of "Type II" confined events have $\Psi_{HFED}/\Phi_{AR} \geqslant 1.0 \times 10^{-21}$ degree $Mx^{-1}$. The flare reconnection flux $\Phi_{RIB}$ and the FWHM duration of the flares $\tau_{FWHM}$ both give larger log-mean values for "Type I" than "Type II" flares.

For "Type I" confined flares, the slipping motions of flare loops and ribbon substructures are the signatures of slipping reconnections along the QSLs (Aulanier et al. 2012; Janvier et al. 2013). Flipping or slipping of magnetic fields was predicted to be a property of all 3D reconnection models (Priest & Forbes 1992; Li et al. 2021b), due to the continuous change of field line connections during their passage through a diffusion region. The slipping motions of flare loops and ribbon substructures have been reported during eruptive flares (Li & Zhang 2015; Dudík et al. 2016), which can be interpreted by the standard flare model in 3D (Janvier et al. 2015). However, recent studies showed that the slipping motions can also be observed during confined flares (Li et al. 2018, 2019). They cannot be simply explained by the 3D standard flare picture, which requires the presence of an eruptive flux rope. Moreover, unlike what the standard flare model in 2D/3D predicts, the PFLs in "Type I" flares are formed above the stable filament, and the flare does not involve the magnetic field of the filaments. The appearance of this type of confined flare is similar to that of the named "atypical" flares in several case studies (Liu et al. 2014; Dalmasse et al. 2015; Joshi et al. 2019). The topological analysis shows that the occurrence of multiple and sequential magnetic reconnections within the complex set of QSLs led to the observed confined flare (Liu et al. 2014; Dalmasse et al. 2015). Li et al. (2019) suggested that the magnetic configuration of "Type I" flares is very complex with two or more QSLs overlying the core magnetic structure, and multiple slipping reconnections along these QSLs results in the occurrence of the flare. Numerical magnetohydrodynamic (MHD) modeling presents that a tether-cutting reconnection between the sheared magnetic arcades leads to "Type I" flares (Jiang et al. 2016).

"Type II" confined flares can be simply explained by the standard flare model in 2D. The flare is caused by the failed eruption of a filament, and a current sheet forms in the corona, right below the erupting filament. The MHD simulation results showed that the energy of the twist flux rope is insufficient to break through the overlying field, whose lines form a confining cage, but its twist is large enough to trigger a kink instability, leading to the confined flare (Török & Kliem 2005; Amari et al. 2018). There is another possibility that the eruptive structure first enters the local torus-unstable region, and then keeps rising and enters the torus-stable region when the decay index of overlying magnetic fields presents a saddle-like profile (Guo et al. 2010; Baumgartner et al. 2018; Luo & Liu 2022). It was suggested that the initiation and development of these types of confined flares are similar to that of eruptive flares, and the main difference is the confinement of the background field (Huang et al. 2020). The overlying large-scale coronal arcades observed during the flares probably correspond to the restraining arcades above the flaring region.

The measures of AR magnetic parameters show that the ARs generating "Type I" flares have a larger $\Phi_{AR}$, with the average of the log values being $8.8 \times 10^{22}$ Mx, much larger than that of "Type II" flares ($5.1 \times 10^{22}$ Mx). It has been revealed that $\Phi_{AR}$ describes the background field confinement overlying the flaring region, which is supported by the high positive correlation between the critical decay index height and $\Phi_{AR}$ (Li et al. 2020, 2021a). The higher $\Phi_{AR}$ of "Type I" flares indicates that the constraints of the overlying field for "Type I" flares are much stronger than "Type II" flares. The relative nonpotentiality parameter $\Psi_{HFED}/\Phi_{AR}$ can provide a good ability for distinguishing the two types of confined flares. It needs to be noted that the relative parameter does not seem to be an entirely clear predictor of the type of flares, considering that there is a certain degree of overlap between the two types of flares. The relative parameter $\Psi_{HFED}/\Phi_{AR}$ probably indicates the balance between the upward force that drives the eruptions and the downward force that suppresses the eruptions, which is an important parameter in determining the capability of ARs to produce eruptive flares (Li et al. 2022). We suggest that for "Type I" confined flares the overlying magnetic field may be too strong to allow the filament to erupt. The other reason explaining the stable filament during "Type I" flares is probably the small twist values of the filament, impossible to trigger the kink instability (Li et al. 2019). We also find that "Type I" and "Type II" confined flares significantly differ regarding the reconnection flux $\Phi_{RIB}$ and the flare duration $\tau_{FWHM}$, with "Type I" cases having larger log-mean values. The moderate correlation was obtained for $\Phi_{RIB}$ versus $\tau_{FWHM}$, consistent with the results of Toriumi et al. (2017). Their relation implies that the reconnection processes continue for longer when more magnetic flux is involved.

Our statistical study shows that the "Type I" confined flares are numerous and occupy as high as 40% of all the large confined flares, which is pointed out for the first time to our knowledge. "Type I" confined flares cannot be interpreted by the standard flare model in 2D and its extension in 3D. The true distinction between "Type I" and "Type II" confined flares might be the presence or absence of flux ropes. In "Type I" flares, the stable filaments probably correspond to sheared magnetic arcades, not flux ropes (Li et al. 2019). "Type II" flares are generated due to the failed eruption of flux ropes. In future, we can carry out the extrapolations of coronal magnetic fields for the analyzed database and give a definite answer. Moreover, the triggering mechanism and the development process of "Type I" confined flares are still unknown, so further topological analyses and numerical simulations of these types of flares are required to build the 3D MHD flare models.

We are grateful to Dr. Jaroslav Dudik for useful suggestions. This work is supported by the B-type Strategic Priority Program of the Chinese Academy of Sciences (XDB41000000), the National Key R&D Program of China (2019YFA0405000), the National Natural Science Foundations of China (11903050, 12073001, 11790304, 11873059, and 11790300), Key Programs of the Chinese Academy of Sciences (QYZDJ-SSW-SLH050), Yunnan Academician Workstation of Wang Jingxiu (No. 202005AF150025) and NAOC Nebula Talents Program. SDO is a mission of NASA's Living With a Star Program.

## ORCID iDs

Xuchun Duan  https://orcid.org/0000-0001-6655-1743
Ting Li  https://orcid.org/0000-0001-6655-1743

## References

Amari, T., Canou, A., Aly, J.-J., Delyon, F., & Alauzet, F. 2018, Natur, 554, 211
Aulanier, G., Janvier, M., & Schmieder, B. 2012, A&A, 543, A110
Aulanier, G., Pariat, E., Démoulin, P., et al. 2006, SoPh, 238, 347
Baumgartner, C., Thalmann, J. K., & Veronig, A. M. 2018, ApJ, 853, 105
Bobra, M. G., Sun, X., Hoeksema, J. T., et al. 2014, SoPh, 289, 3549
Carmichael, H. 1964, NASSP, 50, 451
Chen, A. Q., & Wang, J. X. 2012, A&A, 543, A49
Chen, H., Yang, J., Ji, K., et al. 2019, ApJ, 887, 118
Cui, Y., Wang, H., Xu, Y., et al. 2018, JGRA, 123, 1704
Dalmasse, K., Chandra, R., Schmieder, B., et al. 2015, A&A, 574, A37
Dudík, J., Janvier, M., Aulanier, G., et al. 2014, ApJ, 784, 144
Dudík, J., Polito, V., Janvier, M., et al. 2016, ApJ, 823, 41
Forbes, T. G. 2000, JGR, 105, 23153
Guo, Y., Ding, M. D., Schmieder, B., et al. 2010, ApJL, 725, L38
Gupta, M., Thalmann, J. K., & Veronig, A. M. 2021, A&A, 653, A69
Hirayama, T. 1974, SoPh, 34, 323
Huang, Z. W., Cheng, X., & Ding, M. D. 2020, ApJL, 904, L2
Janvier, M., Aulanier, G., & Démoulin, P. 2015, SoPh, 290, 3425
Janvier, M., Aulanier, G., Pariat, E., et al. 2013, A&A, 555, A77
Ji, H., Wang, H., Schmahl, E. J., et al. 2003, ApJL, 595, L135
Jiang, C., Wu, S. T., Yurchyshyn, V., et al. 2016, ApJ, 828, 62
Joshi, N. C., Zhu, X., Schmieder, B., et al. 2019, ApJ, 871, 165
Kazachenko, M. D., Lynch, B. J., Savcheva, A., et al. 2022, ApJ, 926, 56
Kazachenko, M. D., Lynch, B. J., Welsch, B. T., et al. 2017, ApJ, 845, 49
Kopp, R. A., & Pneuman, G. W. 1976, SoPh, 50, 85
Lemen, J. R., Title, A. M., Akin, D. J., et al. 2012, SoPh, 275, 17
Li, D., Ning, Z. J., & Zhang, Q. M. 2015, ApJ, 807, 72
Li, T., Chen, A., Hou, Y., et al. 2021a, ApJL, 917, L29
Li, T., Hou, Y., Yang, S., et al. 2018, ApJ, 869, 172
Li, T., Hou, Y., Yang, S., et al. 2020, ApJ, 900, 128
Li, T., Liu, L., Hou, Y., et al. 2019, ApJ, 881, 151
Li, T., Priest, E., & Guo, R. 2021b, RSPSA, 477, 20200949
Li, T., Sun, X., Hou, Y., et al. 2022, ApJL, 926, L14
Li, T., & Zhang, J. 2014, ApJL, 791, L13
Li, T., & Zhang, J. 2015, ApJL, 804, L8
Liu, R., Titov, V. S., Gou, T., et al. 2014, ApJ, 790, 8
Luo, R., & Liu, R. 2022, ApJ, 929, 2
Masson, S., Pariat, E., Aulanier, G., et al. 2009, ApJ, 700, 559
Nindos, A., & Andrews, M. D. 2004, ApJL, 616, L175
Pesnell, W. D., Thompson, B. J., & Chamberlin, P. C. 2012, SoPh, 275, 3
Priest, E. R., & Démoulin, P. 1995, JGR, 100, 23443
Priest, E. R., & Forbes, T. G. 1992, JGR, 97, 1521
Savcheva, A., Pariat, E., McKillop, S., et al. 2015, ApJ, 810, 96
Scherrer, P. H., Schou, J., Bush, R. I., et al. 2012, SoPh, 275, 207

Shibata, K., & Magara, T. 2011, LRSP, 8, 6
Sturrock, P. A. 1966, Natur, 211, 695
Sun, X. 2013, arXiv:1309.2392
Sun, X., Bobra, M. G., Hoeksema, J. T., et al. 2015, ApJL, 804, L28
Sun, X., Hoeksema, J. T., Liu, Y., et al. 2013, ApJ, 778, 139
Svestka, Z., & Cliver, E. W. 1992, in IAU Colloq. 133, Eruptive Solar Flares, ed. Z. Svestka, B. V. Jackson, & M. E. Machado (New York: Springer), 1
Toriumi, S., Schrijver, C. J., Harra, L. K., et al. 2017, ApJ, 834, 56
Török, T., & Kliem, B. 2005, ApJL, 630, L97
Wang, D., Liu, R., Wang, Y., et al. 2017, ApJL, 843, L9
Wang, Y., & Zhang, J. 2007, ApJ, 665, 1428
Yang, K., Guo, Y., & Ding, M. D. 2015, ApJ, 806, 171
Yang, S., Zhang, J., & Xiang, Y. 2014, ApJL, 793, L28
Yashiro, S., Akiyama, S., Gopalswamy, N., et al. 2006, ApJL, 650, L143
Zemanová, A., Dudík, J., Aulanier, G., et al. 2019, ApJ, 883, 96
Zhang, Y., Zhang, Q., Song, D., et al. 2022, ApJS, 260, 19
Zhao, J., Gilchrist, S. A., Aulanier, G., et al. 2016, ApJ, 823, 62